\documentclass[structabstract]{aa}
\usepackage{graphicx}
\usepackage{txfonts}
\begin{document}
   \title{Generation of Electrostatic Waves via Parametric Instability and
Heating of Solar Corona}


\author{G. Machabeli \inst{1}
  \and G.Dalakishvili \inst{1,2}
    \and V. Krasnoselskikh \inst{3}
}

   \institute{
  Institute of Theoretical Physics, Ilia State University,
Kakutsa Cholokashvili Ave 3/5 Tbilisi 0162, GEORGIA
  \and
  Faculty of Engeneering, Ilia State University,Kakutsa
Cholokashvili Ave 3/5 Tbilisi 0162, GEORGIA
  \and
 CNRS-University of Orl´eans 3A Avenue de la Recherche Scientifique
45071 Orl´eans CEDEX 2 FRANCE\\
  }



  \abstract
{In the upper layers of the solar atmosphere the temperature
increases sharply. We studied possibility of the transfer of
neutrals motion energy into the electrostatic waves.Electrostatic
waves could damp in the upper layers of the solar atmosphere and
their energy could be transformed into the thermal energy of the
solar atmosphere plasma.}
{We have used the two fluid approximation when studying plasma
dynamics in the low altitudes of the solar atmosphere. In order to
study evolution of disturbances of high amplitudes the parametric
resonance technique is used.}
{ The dispersion relation for the electrostatic waves excited due
tot he motion of neutrals is derived. The frequencies of
electromagnetic waves which could be excited due to existence of the
acoustic wave are found. The increment of excited electrostatic
waves are determined.}
{The motion of the neutrals in the lower solar atmosphere, where
ionization rate is low, could excite electrostatic waves. In the
upper solar atmosphere the ionization rate increases and motion of
the neutrals could not support electrostatic waves and these waves
could damp due to the collision of the charged particles. The energy
of the damping waves could be transformed into the thermal energy of
the plasma in the upper atmosphere.}




   \maketitle
%

\section{Introduction}

In the solar atmosphere the temperature raises sharply from
photospheric 6000K to few $10^{6}K$ in the corona. The problem of
heating solar atmosphere is a long standing problem in solar
Physics. The detailed physical mechanism of coronal heating is not
yet well understood, fundamental problems have to be solved in order
to find proper theoretical models explaining observational data.
Fundamental questions related with the coronal heating include
identification of the source of thermal energy, proposition of the
 energy release mechanism and comparison of theoretically
predicted values of the physical characteristics with observations
(Klimchuk, 2006). Investigation of the coronal heating needs the
study of the highly disparate spatial scales, in particular the
study of  MHD wave dispassion, MHD reconnection, quasi-periodic
oscillations and small-scale bursts occurring in corona is necessary
(Klimchuk, 2006, Walsh \& Irelan, 2003).In addition, the heating of
the chromosphere requires much more energy, and this related process
is equally difficult to understand.

Based on the UV and X-ray observations of the sun, Parker suggested
(1988) that the X-ray solar corona is created by the dissipation at
the many tangential discontinuities arising spontaneously in the
bipolar fields of the solar active regions as consequence of random
continuous motion of the footpoints of the field in the photospheric
convection. Parker named these elementary energy release vents as
nanoflares. Number of studies were dedicated to finding
observational evidences of nanoflares (Cargill, 1994, Sakamoto et
al., 2008, Vekstein, 2009). Prominent researchers found Parkers's
idea very promising and investigated  nanoflares contribution into
energy budget of the solar corona (Klimchuk, 2006).

Parker's model does not address following questions: where the
excess magnetic energy comes form and what is its original source?
According to Rappazzo et al., (2008) slow photospheric motions
induce a Poynting flux which saturates by driving an anisotropic
turbulent cascade dominated by magnetic energy. In this case
magnetic field lines are barely tangential and current sheets are
continuously formed and dissipated (current sheets correspond to
tangential discontinuities Proposed by Parker).

Another mechanism was suggested by Krasnoselskikh et al., (2010),
that electric waves could be generated by motion of neutral gas. A
huge energy reservoir exists in the form of turbulent motion of
neutral gas at end beneath the photosphere. It is widely accepted
that this energy could be partially transformed into the excess
magnetic energy in the chromosphere and corona.

The mechanism of transformation of neutral gas motion energy into
the electromagnetic disturbances was investigated by Aburjania \&
Machabeli (1998). In this paper authors derived dispersion equations
for coupled acoustic and electromagnetic modes. The authors
considered studied parametric instability. This technique enables to
study nonlinear processes, i.e. pumping energy from the acoustic
waves with high amplitudes. In Aburjania \& Machabeli (1998) authors
implied discussed mechanism to explain observed data and predict
characteristic  expected atmospheric wave preprocesses before the
earthquakes.

In order to study excitation of the electromagnetic waves in solar
atmosphere we use data, describing parameters of solar atmosphere
given by Vernaza et al. (1981). We consider the case where excited
velocities, electric field and  wave vector and  are parallel to
external magnetic field.


\section{Main consideration}
In the low layers of the solar atmosphere (up to 1000 km, see
Vernaza et al. 1981), the concentration of neutral is much higher
than concentration of the charged component of plasma. At this
altitudes of the solar atmosphere the motion of neutrals could
influence dynamics of electrons and ions, therefore the energy of
motion of neutrals could be transformed into the electromagnetic
energy, in particular energy of the acoustic waves could be pumped
into the energy of the electrostatic waves. The intensive acoustic
wave draws charged particles into the motion due to  friction with
neutrals.

Here we investigate how the energy of acoustic waves with high
amplitude, could be transformed into the energy of electrostatic
waves. When considering the disturbances of high amplitudes, we
could not neglect nonlinear terms, more over these terms play a key
role in the process. The nonlinear pumping of the electromagnetic
wave energy into the plasma, due to the parametric instability was
studied by Silin (1973), in order to investigate neutral wave energy
pumping into the plasma the parametric instability mechanism was
studied by Aburjania Machabeli (1998), in this work we also apply
parametric instability technique.

 If we write and sum the equations of motion for electrons and
ions,

 we will get:

$$
m_{e}n_{e}\frac{d\upsilon_{e}}{dt}+m_{i}n_{i}\frac{d\upsilon_{i}}{dt}=
$$
$$
-\nabla\left(p_{e}+p_{i}\right)-m_{i}n_{i}\nu_{in}\left(\upsilon_{i}-\upsilon_{n}\right)+
m_{i}n_{i}\nu_{ie}\left(\upsilon_{e}-\upsilon_{i}\right)
$$
$$
+m_{e}n_{e}\nu_{ei}\left(\upsilon_{i}-\upsilon_{e}\right)+m_{e}n_{e}\nu_{en}\left(\upsilon_{n}-\upsilon_{e}\right).\eqno(1)
$$

Here is assumed that charged particle move along the magnetic field
lines ant the direction of electric field and wave vector coincides
with the  direction of the magnetic field, i.e we consider
electrostatic waves.

In the low altitude (beyond 1000km) of the solar atmosphere the
relations given below by Eq.~2a-2d are satisfied:

$$
\frac{\nu_{en}}{\nu_{in}}=\sqrt{\frac{m_{i}}{m_{e}}},\eqno(2a)
$$

$$
\frac{\nu_{ei}}{\nu_{ie}}=\sqrt{\frac{m_{i}}{m_{e}}},\eqno(2b)
$$

$$
\frac{\nu_{in}}{\nu_{ni}}\approx\frac{n_{0n}}{n_{0i}}\gg 1,\eqno(2c)
$$

If the electric field changes with frequency much less than the
cyclotron frequency we could assume that the charged particle moves
in a constant electric field during the Larmour rotation period. In
this case the velocity of the charged particle is: $\upsilon_{\alpha
z}=q_{\alpha}\frac{E_{z}}{m_{\alpha}}t$ (here indices $\alpha$ shows
the sort of particle and $t$ is time). In order to define the
interval between collisions of the charged particle kind $\alpha$
with neutrals we use the law of statistical distribution (see
Aburjania \& Machabeli 1998), i.e the probability that collision of
particle $\alpha$ with neutrals happens between $t$ and $t+dt$ is:
$f_{\alpha n}=\nu_{\alpha n}e^{-\nu_{\alpha n}t}dt.$ Multiplication
of $\upsilon_{\alpha z}=q_{\alpha}\frac{E_{z}}{m_{\alpha}}t$ by
$f_{\alpha n}=\nu_{\alpha n}e^{-\nu_{\alpha n}t}dt$ and integration
by time leads to the following relation between velocities of
electron and ions:

$$
\upsilon_{ez}=-\beta\upsilon_{iz},\eqno(2d)
$$

here:

$\beta=\frac{m_{i}\nu_{in}}{m_{e}\nu_{en}}.$ Eq.~2d helps us to
solve us three component hydrodynamic equations system.

After taking into account relations given by Eq.~2a-2d and comparing
similar terms in Eq.~(1), we obtain:
$$
\frac{m_{i}n_{i}}{m_{e}n_{e}}\frac{d\upsilon_{i}/dt}{d\upsilon_{e}/dt}\approx\frac{n_{i}\omega_{pi}}{n_{e}\omega_{pe}}\approx\sqrt{\frac{m_{i}}{m_{e}}}\gg
1,\eqno(3a)
$$

$$
\frac{m_{i}n_{i}\nu_{in}}{m_{e}n_{e}\nu_{en}}=\sqrt{\frac{m_{i}}{m_{e}}}\gg
1,\eqno(3b)
$$

$$
\frac{m_{i}n_{i}\nu_{ie}(\upsilon_{e}-\upsilon_{i})}{m_{e}n_{e}\nu_{ei}(\upsilon_{e}-\upsilon_{i})}=\sqrt{\frac{m_{i}}{m_{e}}}\gg
1,\eqno(3c)
$$

$$\upsilon_{e}\gg\upsilon_{i}.\eqno(3d)$$

After taking into account Eq.~(3a-3d) and neglecting electron
inertia, we could rewrite Eq.~1 as follows:
$$
\frac{\partial\upsilon_{i}}{\partial
t}+\left(\nabla\upsilon_{i}\right)\upsilon_{i}=
$$
$$
-\frac{1}{m_{i}n_{i}}\nabla\left(p_{e}+p_{i}\right)-\nu_{in}\left(\upsilon_{i}-\upsilon_{n}\right)+\nu_{ie}\left(\upsilon_{e}-\upsilon_{i}\right).\eqno(4)
$$

If we represent ion velocity as:
$$
\upsilon_{i}-\upsilon_{n}=-V_{i},\eqno(5a)
$$

we express the velocity of neutrals as follows:

$$
\upsilon_{n}=V_{0}\cos(\Omega t+\Psi),\eqno(5b)
$$

Eq.~5b describes oscillatory motion of the neutrals, this type of
motion accompanies the excitation and propagation of waves
interacting with neutrals. In Eq.~5b  $\Psi$ is casual, initial
phase of neutrals oscillations. Because of the collisions between
particles the phase of the oscillations changes casually, the final
expressions of the variables obtained after solving hydrodynamic
equations should be averaged by $\Psi$.

After substitution of (5a-5b) in Eq.~(4)
 we get

$$
\frac{\partial V_{i}}{\partial t}-V_{0}\Omega\sin(\Omega
t+\Psi)+ikV_{i}V_{0}\cos(\Omega t+\Psi)=
$$
$$
-\frac{1}{m_{i}n_{i}}\nabla\left(p_{e}+p_{i}\right)-V_{i}\left(\nu_{in}+\beta\nu_{ie}\right)-\nu_{ie}\beta
V_{0}\cos(\Omega t+\Psi),\eqno(6a)
$$

and
$$
\frac{\partial V_{i}}{\partial t}-V_{0}\left[\Omega\sin(\Omega
t+\Psi)+\nu_{ei}\cos(\Omega t+\Psi)\right]+ikV_{i}V_{0}\cos(\Omega
t+\Psi)=
$$
$$
-\frac{1}{m_{i}n_{i}}\nabla\left(p_{e}+p_{i}\right)-V_{i}\left(\nu_{in}+\nu_{ei}\right).\eqno(6b)
$$

It is convenient to use notation:
$$
V_{i}\equiv U\exp\left[i a\sin(\Omega t+\Psi)\right].\eqno(7)
$$

Substitution of Eq.~(7) into Eq.~(6b) leads to equation:

$$
\frac{\partial U}{\partial t}-V_{0}\left[\Omega\sin(\Omega
t+\Psi)+\nu_{ei}\cos(\Omega t+\Psi)\right]\exp\left[-i a\sin(\Omega
t+\Psi)\right]=
$$
$$
-\frac{1}{m_{i}n_{i}}\nabla\left(p_{e}+p_{i}\right)\exp\left[-i
a\sin(\Omega t+\Psi)\right]-U\left(\nu_{in}+\nu_{ei}\right).\eqno(8)
$$

Let us write equation of continuity:

$$
\frac{\partial{(n_{0}+n_{e})}}{\partial
t}+\nabla(n_{0}+n_{e})\mathbf{\upsilon_{e}}=0,\eqno(9a)
$$

unperturbed value of concentration-$n_{0}$ does not depend on time,
therefore $\frac{\partial n_{0}}{\partial t}=0.$ It is convenient to
introduce notation $n_{e}=N_{e}(t)\exp\left[-ib\sin(\Omega
t+\Psi)\right],$ here $a=\frac{kV_{0}}{\Omega},
 b=\beta\frac{kV_{0}}{\Omega}, \beta=\sqrt{\frac{m_{i}}{m_e}}.$ For
 $N_{e}$ we get equation:

 $$
 \frac{\partial N_{e}}{\partial t}-ik\beta
 U(t)n_{0}\exp\left[-ib\sin(\Omega t+\Psi)+ia\sin(\Omega
 t+\Psi)\right],\eqno(9b)
 $$

solution of Eq.~9b gives

$$
n_{e}=ikn_{0}\beta\exp[ib\sin(\Omega t+\Psi)]\times
$$
$$
\times\sum
J_{\eta}(b)\exp(i\eta\Psi)\int\frac{U(\omega')\exp\left[-i(\omega'+\eta\Omega)\right]}{\omega'+\eta\Omega}d\omega'.\eqno(9c)
$$

For pressure gradients we have:

$$
\frac{1}{m_{i}n_{0}}\nabla
p_{i}=\frac{k^{2}V_{Ti}^{2}}{\omega},\eqno(10a)
$$

$$
\frac{1}{m_{i}n_{0}}\nabla
p_{e}=\frac{-ikT_{e}n_{e}}{m_{i}n_{0}},\eqno(10b)
$$
here $V_{Ti}=\sqrt{\frac{\gamma k T_{i}}{m_{i}}}.$

After substitution of expressions (9c), (10a), (10b) into equation
(8) and averaging by $\Psi$ (it is casual phase of neutrals
oscillations) we could derive equation for frequency of excited
electrostatic waves:

$$
\omega^{2}+i\nu\omega-k^{2}V_{Ti}^{2}-\beta(kV_{se})^{2}\omega\sum{\frac{J_{j_{0}}^{2}(b)}{\omega-j_{0}\Omega}}=0,\eqno(11)
$$

where:

$\nu=\nu_{in}+\nu_{ei},$ $\omega_{0}=j_{0}\Omega,$

$V_{se}=\sqrt{\frac{\gamma k T_{e}}{m_{i}}}.$

For $\nu_{in}$ and $\nu_{ei}$ the expressions given in Braginskii
(1963) are used.

Because of the nonlinear character of the investigated processes,
the averaging by casual phase $\Psi$, the terms of hydrodynamic
equations, which depend on $\Psi$ do not vanish.

 In Eq.~9 we see that resonance occurs when $\omega\approx
j_{0}\Omega$. It is convenient to represent $\omega$ in the
following way:
$$
\omega=\omega_{0}+i\delta.\eqno(12)
$$
here $\omega_{0}=j_{0}\Omega$

after substitution of  Eq.~12 into Eq.~11 we obtain expression for
frequency of the excited wave and their increment:

$$
\omega_0=\pm\sqrt{\delta\nu+k^{2}V_{Ti}^{2}+J_{j_{0}}^{2}(b)\beta(kV_{se})^{2}},\eqno(13a)
$$
$$
\delta=-\frac{\nu}{4}\pm\sqrt{\frac{\nu^{2}}{16}+0.5J_{j_{0}}^{2}(b)\beta(kV_{se})^{2}},\eqno(13b)
$$

We could solve equations 13a and 13b numerically.  Fig.~1 shows
values of the function $f_{\omega}(k,j_{0})$, where
$f_{\omega}(k,j_{0})=\omega$ if the Eqs.13a-b have solution and
$f_{\omega}(k,j_{0})=0$ if Eqs.13a-b do not have solutions. Fig.~2
represents values of the function  $f_{\delta}(k,j_{0})$, where
$f_{\delta}(k,j_{0})=\delta$ if the Eqs.13a-b have solution and
$f_{\delta}(k,j_{0})=0$ if Eqs.13a-b do not have solutions. When
solving equations the parameters, given in paper by Vernazza et al.
(1981), are used.

In our model the only source of excitation of electrostatic waves is
the oscillation of neutrals, in the case of absence of acoustic
waves i.e. $V_{0}=0$, in the fourth term of Eq.~11
$\sum{\frac{J_{j_{0}}^{2}(b)}{\omega-j_{0}\Omega}}=\frac{J_{0}(0)}{\omega}=\frac{1}{\omega}$,
therefore we have:

$$
\omega^{2}+i\nu_{ei}\omega-k^{2}V_{Ti}^{2}-\beta(kV_{se})^{2}=0,\eqno(14)
$$

the solution of Eq.~14 is:

$$
\omega=-\frac{i\nu_{ei}}{2}\pm\sqrt{\frac{\nu_{ei}^{2}}{4}
+k^{2}V_{Ti}^{2}+\beta(kV_{se})^{2}}.\eqno(15)
$$

Eq.~15 represent the damping  wave with decrement
$\frac{\nu_{ei}}{2}$ and dispersion relation
$\omega_{0}=\pm\sqrt{\frac{\nu_{ei}^{2}}{4}+k^{2}V_{Ti}^{2}+\beta(kV_{se})^{2}}$,
this result shows the fact that  in  the absence of acoustic waves
the electrostatic waves could not be excited and where the motion of
the neutrals does not play a role the damping of propagating
electrostatic waves will occur.

In the regions of the solar atmosphere where the plasma ionization
rate is high the motion of the neutrals does not influence the
dynamics of the charged particles. Due to collisions between the
charged particles the damping of the electrostatic wave will occur
and the energy of electrostatic waves could be transformed into the
thermal energy of the solar atmosphere. The Fig.~3 shows dependence
of the collision frequencies $\nu_{ei}$ and $\nu_{en}$ on the
altitude in the solar corona, we see that at high altitudes
$\nu_{ei}$  (which is responsible for wave damping) is higher than
$\nu_{en}$. For the Fig.~3 the solar atmosphere model given by
Vernazza et al. (1981) is used. The dependence of the characteristic
frequencies on the altitude in the solar atmosphere is given in
Krasnoselskikh et al. (2010), where the model of the sun described
in Fontenla et al. (2009) is used.

\section{Conclusions}

We investigated possibility of transformation of neutral gas motion
energy into the electromagnetic energy, in the solar atmosphere. We
described pumping of the acoustic wave energy into the energy of
electrostatic waves in the regions of the solar atmosphere where
concentration of the neutrals is much higher than concentration of
the charged particles.

With increasing of the height in the solar atmosphere the ionization
rate of the plasma increases i.e the role of the motion of the
neutrals in the process of wave generation decreases and waves damp
due tot he collision of the charged particles. The energy of the
damping waves could be transferred in to the thermal energy of the
solar atmosphere plasma, at higher altitudes the temperature of the
solar atmosphere plasma increases (Vernazza et al. 1981).

We used two fluid approach and the parametric instability technique
is used.  We have determined that for particular parameters anergy
of acoustic waves of certain frequencies and certain wavelength
could be pumped into the energy of the electrostatic waves. Figs~1
and 2 what kind of acoustic waves could contribute into the growth
of the electrostatic waves.

In this paper the problem was investigated using two fluid MHD
approximation, we find it interesting to study transformation of
neutral gas energy into the electromagnetic energy using kinetic
approach, these approach could lead to more precise results.

\section{Acknowledgments}

G.Machabeli would like to thank A.S. Volokitin for interesting
discussions and he is grateful to LPC2E, CNRS-University of Orl´eans
for support of his research.

\begin{figure*}[t]
\vspace*{1mm}
\begin{center}
\includegraphics[width=10cm]{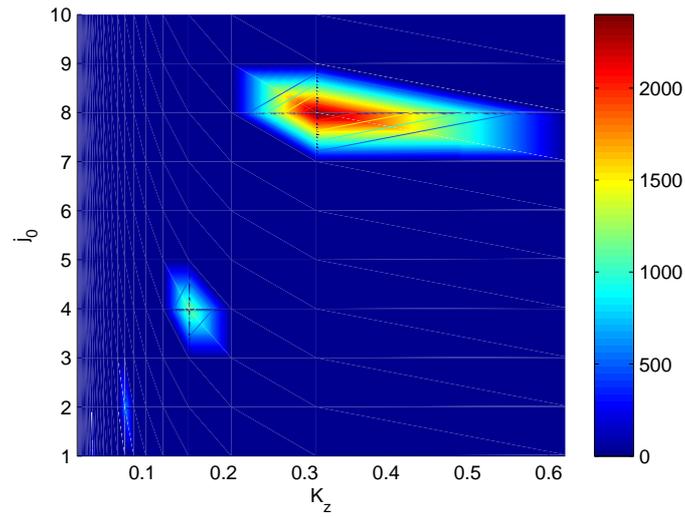}
\end{center} \caption{Dependence of "filtered", excited waves
frequency on wave number and $j_{0}$. Following particular
parameters are chosen: $n_{0n}=10^{21}\rm{m^{-3}},$ $T=4070K$.}
\end{figure*}

\begin{figure*}[t]
\vspace*{1mm}
\begin{center}
\includegraphics[width=10cm]{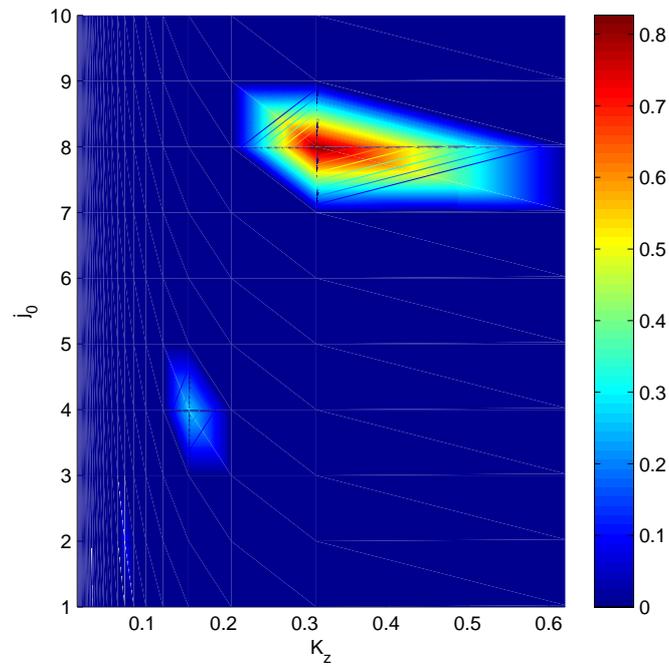}
\end{center} \caption{Dependence of excited waves "filtered"
increment on wave number and $j_{0}$. Following particular
parameters are chosen: $n_{0n}=10^{21}\rm{m^{-3}},$ $T=4070K$. }
\end{figure*}

\begin{figure*}[t]
\vspace*{1mm}
\begin{center}
\includegraphics[width=10cm]{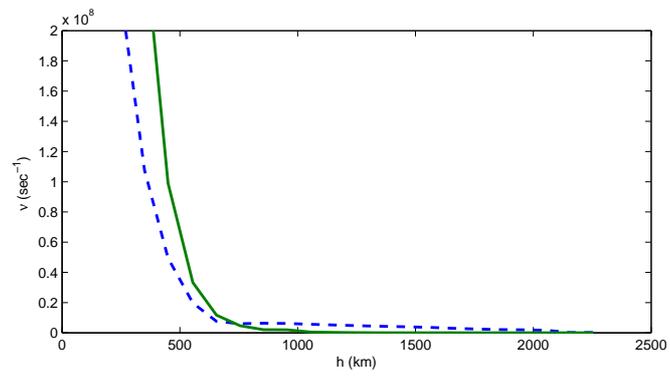}
\end{center} \caption{Dependence of the collision frequencies $\nu_{ie}$-dashed line and $\nu_{en}$-solid line, on the altitude in the solar atmosphere.}
\end{figure*}
\end{document}